\documentclass[twocolumn,showpacs,preprintnumbers,amsmath,prl,amssymb,superscriptaddress]{revtex4-1}

\usepackage[export]{adjustbox}
\usepackage{amsfonts}
\usepackage{amsmath}
\usepackage[makeroom]{cancel}
\usepackage[english]{babel}
\usepackage[T1]{fontenc}
\usepackage{txfontsb}
\usepackage{times}
\usepackage{mathrsfs}
\usepackage{graphicx}% Include figure files
\usepackage{dcolumn}% Align table columns on decimal point
\usepackage{bm}% bold math
\usepackage{wasysym}
\usepackage[colorlinks,bookmarks=true,citecolor=blue,linkcolor=red,urlcolor=blue]{hyperref}
\usepackage[tight, FIGTOPCAP, hang, raggedright, nooneline]{subfigure}
\usepackage{hyperref}
\usepackage{xcolor}
\usepackage{epsfig}
\usepackage{amssymb}
\usepackage{lipsum}

\makeatletter
\def\p@subsection{}
\makeatother

\newcommand{\subfigimg}[3][,]{%
  \setbox1=\hbox{\includegraphics[#1]{#3}}% Store image in box
  \leavevmode\rlap{\usebox1}% Print image
  \rlap{\hspace*{0pt}\raisebox{\dimexpr\ht1+0\baselineskip}{#2}}% Print label
  \phantom{\usebox1}% Insert appropriate spcing
  }
  
 \definecolor{Green}{RGB}{80,182,0}

\usepackage{color}

\newcommand{\la}{\langle}
\newcommand{\ra}{\rangle}

\renewcommand{\Re}{\operatorname{Re}}

\graphicspath{./Images/}
\usepackage{epstopdf}

\begin{document}
\title{Logarithmic spreading of out-of-time-ordered correlators without many-body localization}
\author{Adam Smith}
\email{as2457@cam.ac.uk}
\affiliation{T.C.M. group, Cavendish Laboratory, J.~J.~Thomson Avenue, Cambridge, CB3 0HE, United Kingdom}
\author{Johannes Knolle}
\affiliation{T.C.M. group, Cavendish Laboratory, J.~J.~Thomson Avenue, Cambridge, CB3 0HE, United Kingdom}
\affiliation{Blackett Laboratory, Imperial College London, London SW7 2AZ, United Kingdom}
\author{Roderich Moessner}
\affiliation{Max Planck Institute for the Physics of Complex Systems, N\"{o}thnitzer Str. 38, 01187 Dresden, Germany}
\author{Dmitry L.~Kovrizhin}
\affiliation{Rudolf Peierls Centre for Theoretical Physics, 1 Keble Road, Oxford, OX1 3NP, United Kingdom}
\affiliation{NRC Kurchatov institute, 1 Kurchatov sq., 123182, Moscow, Russia}
\date{\today}

\begin{abstract}
Out-of-time-ordered correlators (OTOCs) describe information scrambling under unitary time evolution, and provide a useful probe of the emergence of quantum chaos. Here we calculate OTOCs for a model of disorder-free localization whose exact solubility allows us to study long-time behaviour in large systems. Remarkably, we observe logarithmic spreading of correlations, qualitatively different to both thermalizing and Anderson localized systems. Rather, such behaviour is normally taken as a signature of many-body localization, so that our findings for an essentially non-interacting model are surprising. We provide an explanation for this unusual behaviour, and suggest a novel Loschmidt echo protocol as a probe of correlation spreading. We show that the logarithmic spreading of correlations probed by this protocol is a generic feature of localized systems, with or without interactions.
\end{abstract}

\maketitle

%\textit{Introduction}.~
Out-of-time-ordered correlators have proven to be useful in quantifying the spreading of correlations and entanglement in many-body quantum dynamics. While originally introduced in the context of quasiclassical approaches to quantum systems~\cite{Larkin1969}, recently they have received renewed interest due to their connections with the emergence of quantum chaotic behaviour~\cite{Berry1989,Maldacena2016,Bohrdt2017}. This has generated a flurry of activity on OTOCs in the studies of entanglement and information scrambling in integrable~\cite{Lin2018,Dora2017,McGinley2018}, thermalizing~\cite{Bohrdt2017} and many-body-localized (MBL) systems~\cite{Chen2017,Huang2017,Fan2017}, where OTOCs have been calculated for a number of important models including the transverse field Ising model~\cite{Lin2018}, Luttinger-liquids~\cite{Dora2017}, and random unitary circuits~\cite{Nahum2017a,VonKeyserlingk2017,Rakovszky2017,Zhou2018}. Beyond these examples, the calculation of OTOCs is a difficult task. Nonetheless, some generic features of OTOCs are emerging.

We consider OTOCs $C(t) = \la | [ \hat{A}(t),\hat{B} ] |^2 \ra/2$
where $\hat{A},\hat{B}$ are two local operators in their Heisenberg representation, and in the following we assume that expectation values are taken with respect to pure quantum states. For local operators, the infinite temperature OTOC is bounded, having a light-cone causality structure
and exponentially suppressed correlations outside the light-cone~\cite{Lieb1972}. However, in many paradigmatic models the behaviour differs. For example, in generic thermalizing systems, and for random unitary circuits, %~\cite{Nahum2017a,VonKeyserlingk2017,Rakovszky2017,Zhou2018} 
the OTOC saturates to a non-zero value within the light-cone~\cite{Bohrdt2017}, whereas for particular spin components for the transverse-field Ising model the OTOC decays back to zero~\cite{Lin2018}.

Localized systems provide a distinct setting where correlations do not spread linearly. For example, in non-interacting disordered systems correlations (including OTOCs) don't spread beyond the localization length~\cite{Chen2017,Huang2017}, while in many-body localized systems they extend beyond the localization length, albeit logarithmically slowly~\cite{Znidaric2008,Bardason2012,Huang2017}.

Here we show that the situation in disordered systems is even more complex: in particular, many-body localization is \emph{not} a necessary prerequisite for logarithmic OTOC spreading. We demonstrate this in a disorder-free localization model that we have introduced recently~\cite{Smith2017,Smith2017_2,Smith2018,Smith2018_2}, which consists of spinless fermions coupled via a minimal coupling to dynamical Ising spins.
Here, the OTOCs are accessible to large scale numerics and unambiguously demonstrate spreading of correlations beyond the localization length in an essentially non-interacting localized system. While some of the OTOCs for the fermions corresponds to those discussed previously in the context of Anderson localization~\cite{Chen2017,Huang2017}, the spin degrees of freedom yield a richer set of correlators, which in fact correspond to novel fermion correlators.

The main results of this paper are threefold: (i)~OTOCs in our model can be expressed in terms of a double Loschmidt echo, which can be reduced to disorder-averaged correlators in a model of Anderson localization via a non-linear mapping to free-fermions. (ii)~at short times we find power-law growth of the OTOCs, similar to the behaviour found for integrable models, and in contrast to the exponential growth found in semiclassical and large-$N$ models~\cite{Aleiner2016,Kitaev2015,Maldacena2016a}. (iii)~most remarkably, we observe logarithmic spreading of OTOCs at long times, and we argue that this is a generic feature of localized systems, with or without interactions. Since logarithmic correlation spreading is typically associated with many-body localization, this not only adds a facet to our knowledge of OTOC behavioural types, but also imposes further constraints on their systematic classification.

%%%%%%%%%%%%%%%%%%%%

\textit{Setup of the problem.}~First, we review our model of disorder-free localization~\cite{Smith2017,Smith2017_2,Smith2018}, which is a model of spinless fermions $\hat{f}_j$ living on lattice sites (here we consider a 1D lattice with open boundary conditions) which are minimally coupled to spin-1/2s, $\hat{\sigma}_{j}$, defined on the links between neighbouring sites $j$ and $j+1$. The model is described by the Hamiltonian
\begin{equation}\label{eq: H}
\hat{H} = -\sum_{j=1}^{N-1} J_{j}\hat{\sigma}^z_{j} \left(\hat{f}^\dagger_j \hat{f}^{}_{j+1} + \text{H.c.} \right) - \sum_{j=2}^{N-1} h_j \hat{\sigma}^x_{j-1} \hat{\sigma}^x_{j},
\end{equation}
where $N$ is the number of lattice sites, $J_{j}$ and $h_j$ are the fermion tunnelling amplitude and spin-couplings correspondingly, which we assume to be position independent, $J_{j} = J$ and $h_j = h$. 

The Hamiltonian Eq.~\eqref{eq: H} is an example of a $\mathbb{Z}_2$ lattice gauge theory, closely related to the slave-spin description of the Hubbard model~\cite{Ruegg2010,Zitko2015}. An experimental proposal for simulating a 2D version of the model using current technologies in cold atom experiments was presented in Ref.~\cite{Smith2018_2} and generalization of the model have been been discussed in Refs.~\cite{Smith2018,Prosko2017}. The disorder-free mechanism for localization has also been considered for the case of $U(1)$ gauge fields~\cite{Brenes2018}.

In the following we study OTOCs for the spins,
\begin{equation}\label{eq: OTOC def}
C^{\alpha\beta}_{jl}(t) = \frac{1}{2}\la \Psi | \,| [\hat{\sigma}^\alpha_{j}(t), \hat{\sigma}^\beta_{l} ] |^2 |\Psi\ra = 1 - \Re [F^{\alpha\beta}_{jl}(t)],
\end{equation} 
where $\alpha,\beta \in\{x,z\}$, and $|\Psi\ra$ is an initial state of the fermions and spins, and we defined
\begin{equation}
F^{\alpha\beta}_{jl}(t) = \la \Psi | \hat{\sigma}^\alpha_{j}(t) \hat{\sigma}^\beta_{l} \hat{\sigma}^\alpha_{j}(t) \hat{\sigma}^\beta_{l} |\Psi\ra.
\end{equation}
We take the initial state to be a tensor product $|\Psi\ra = |S\ra\otimes |\psi\ra$ of the spins polarized along the $z$-axis, and a Slater determinant for the fermions describing a half-filled Fermi-sea. Thus the OTOC in Eq.~\eqref{eq: OTOC def} corresponds to a global quantum quench~\cite{Essler2016,Vasseur2016}. This initial state is also regularly prepared in cold atom optical lattice experiments~\cite{Schneider2008,Hackermuller2010,Schneider2012}. Note that both the Hamiltonian and the initial state are translationally invariant.

%%%%%%%%%%%%%%%%%%%%%%%%%%%%%%%%

\textit{Double Loschmidt echo.}~It is instructive to rewrite the correlator $F^{\alpha\beta}_{jl}(t)$ in terms of an average that is similar to a Loschmidt echo. Using standard commutation relations for the spin-operators we have $\hat{\sigma}^\alpha_{j} e^{\pm i \hat{H} t} = e^{\pm i \hat{H}^{\alpha}_{j} t} \hat{\sigma}^\alpha_{j}$ and $\hat{\sigma}^\alpha_{j}\hat{\sigma}^\beta_{l} e^{\pm i \hat{H} t} = e^{\pm i \hat{H}^{\alpha\beta}_{jl} t} \hat{\sigma}^\alpha_{j}\hat{\sigma}^\beta_{l}$, where $\hat{H}^\alpha_{j}$ is the Hamiltonian with redefined couplings $h_{j},h_{j+1} \rightarrow -h_{j},-h_{j+1}$ for $\alpha = z$, and for $\alpha = x$ we have $J_{j} \rightarrow -J_{j}$, and similarly for $\hat{H}^{\alpha\beta}_{jl}$. Hence, we can write the correlator in terms of a \emph{double} Loschmidt echo
\begin{equation}\label{eq: OTOC F}
F^{\alpha\beta}_{jl}(t) = \la \Psi | e^{i\hat{H}t} e^{-i\hat{H}^\alpha_{j}t} e^{i \hat{H}^{\alpha\beta}_{jl}t} e^{-i\hat{H}^\beta_{l}t} | \Psi\ra.
\end{equation}
%To clarify the shorthand notation used in this expression, let us for concreteness consider $\alpha = x$, $\beta = z$, then $\hat{H}^\alpha_{j}$ has $J_{j} \rightarrow -J_{j}$, $\hat{H}^{\alpha\beta}_{jl}$ has $J_{j},h_l,h_{l+1} \rightarrow -J_{j},-h_l,-h_{l+1}$, and $\hat{H}^\beta_{l}$ has $h_l,h_{l+1} \rightarrow -h_l,-h_{l+1}$, all relative to $\hat{H}$. 
This correlator can be interpreted in a generalized Keldysh formalism with a contour that folds forward and backward in time twice~\cite{Aleiner2016a,Tsuji2017}. The double Loschmidt echo procedure in Eq.~\eqref{eq: OTOC F} measures the spatial influence of local perturbations as a function of time, which is not captured by the standard Loschmidt echo $L(t) = \la\Psi| e^{i(\hat{H}+\hat{V}) t} e^{-i\hat{H}t} |\Psi\ra$, see SM for further interpretation.

%%%%%%%%%%%%%%%%%%%%%%%%%%%%%%%%%%%%

\textit{Free-fermion mapping.}~The calculations of the OTOCs are made possible by an exact mapping to free-fermions~\cite{Smith2018} resulting from the identification of $N-2$ mutually commuting conserved charges $\hat{q}_j = \hat{\sigma}^x_{j-1} \hat{\sigma}^x_{j}(-1)^{\hat{n}_j}$ having eigenvalues $\pm 1$. The mapping proceeds by a duality transformation for the spins~\cite{Fradkin2013} defining new spin degrees of freedom $\hat{\tau}$ on lattice sites via
\begin{equation}
\hat{\tau}^z_j = \hat{\sigma}^x_{j-1} \hat{\sigma}^x_{j}, \qquad \hat{\tau}^x_j \hat{\tau}^x_{j+1} = \hat{\sigma}^z_{j}.
\end{equation}
After introducing new fermions, $\hat{c}^{}_j = \hat{\tau}^x_j \hat{f}^{}_j$, the Hamiltonian acting in each of the charge sectors, labelled by $\{q_j\} = \pm1$, can be written as
\begin{equation}\label{eq: H free fermions}
\hat{H}(q) = -\sum_{j=1}^{N-1} J_{j} (\hat{c}^\dagger_j \hat{c}^{}_{j+1} + \mathrm{h.c.} )+ \sum_{j=2}^{N-1}2 h_j q_j (\hat{c}^\dagger_j \hat{c}^{}_j - 1/2).
\end{equation}
For a given charge configuration this Hamiltonian corresponds to a tight-binding model with a binary potential. 

In terms of conserved charges, our initial state assumes the form~\cite{Smith2017,Smith2018}
\begin{equation}
|\Psi\ra = \frac{1}{\sqrt{2^{N-2}}} \sum_{\{q_j\} = \pm 1} |q_2 q_3\cdots q_{N-1} \ra \otimes |\psi\ra,
\end{equation}
with $|\psi\ra$ the same Slater determinant for $\hat{c}$ fermions as for $\hat{f}$ fermions. Equation~\eqref{eq: OTOC F} in terms of free-fermions reads
\begin{equation}\label{eq: disorder averaged F}
F^{\alpha\beta}_{jl}(t) =
\frac{1}{2^{N-2}} \sum_{\{q_i\} = \pm1} \la \psi | e^{i\hat{H}(q)t} e^{-i\hat{H}^{\alpha}_{j}(q)t} e^{i \hat{H}^{\alpha\beta}_{jl}(q)t} e^{-i\hat{H}^{\beta}_{l}(q)t} | \psi\ra,
\end{equation}
where the sum is over all $2^{N-2}$ charge configurations. This is our first key result, that the spin OTOCs reduce to disorder-averaged \emph{double} Loschmidt echo for free-fermions. Since the Hamiltonian of Eq.~\eqref{eq: H free fermions} is bilinear in fermion operators, the expectation values in every charge sector can be efficiently computed using determinants, as explained in the Appendix of Ref.~\cite{Smith2018}.  We note that the OTOCs for the fermion density operators, $\hat{n}_j = \hat{f}^\dag_j \hat{f}_j$, correspond exactly to disorder-averaged density correlators for the free-fermions, see e.g.~\cite{Huang2017,Chen2017}.

%%%%%%%%%%%%%%%%%%

\begin{figure}[t]
	\centering
	\subfigimg[width=.43\textwidth]{\hspace*{0pt} \textbf{(a)}}{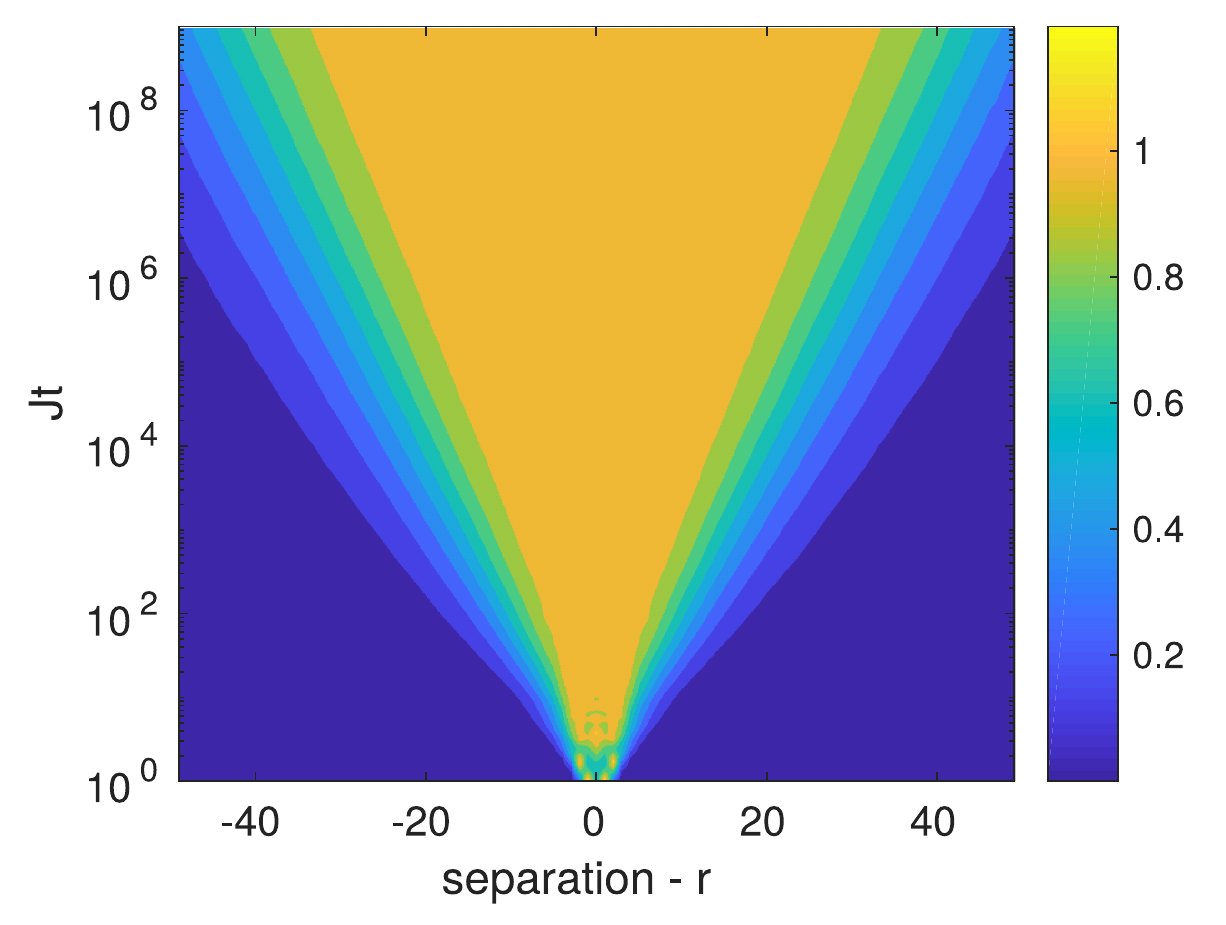}
	\subfigimg[width=.43\textwidth]{\hspace*{0pt} \textbf{(b)}}{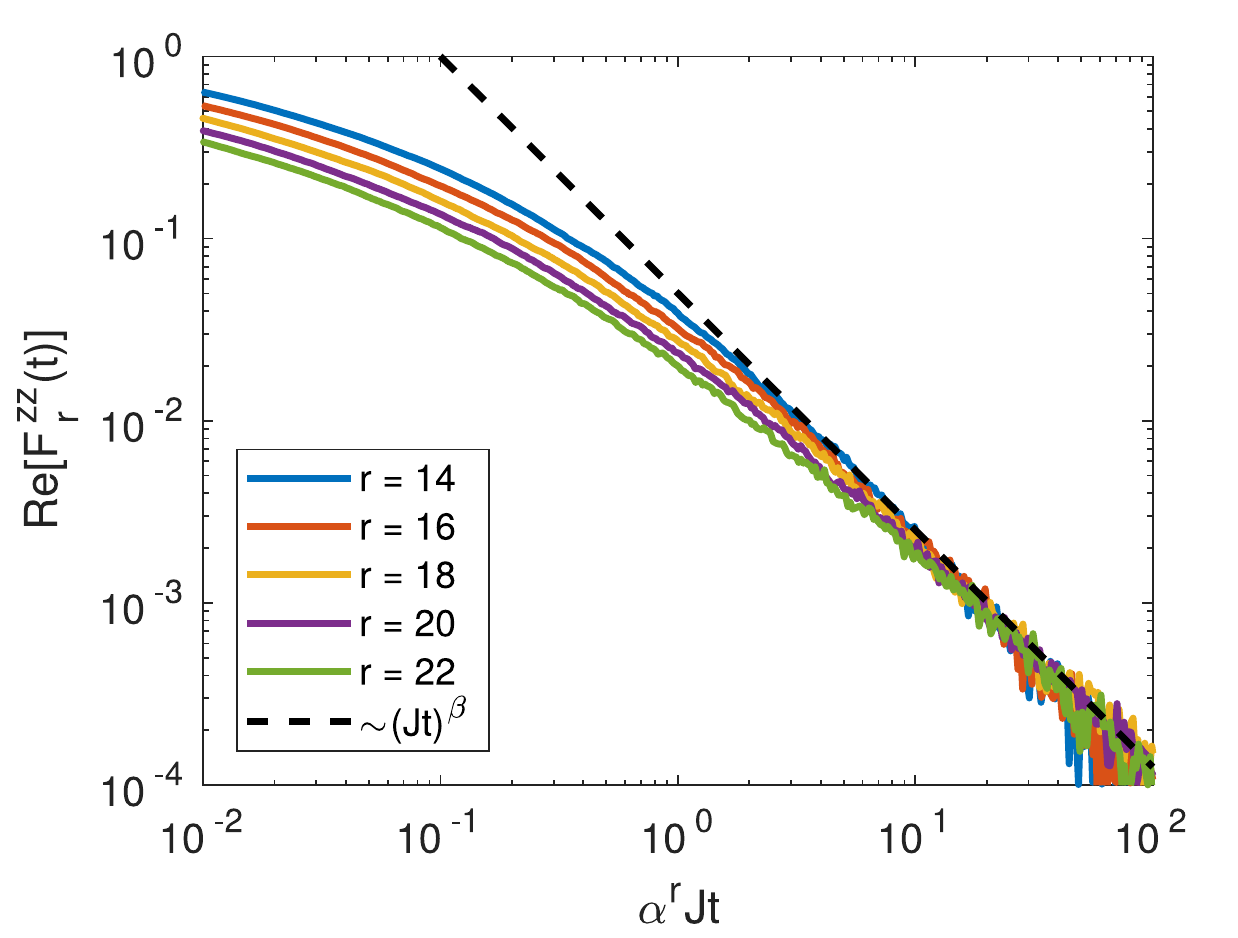}
	\caption{(a) Results for $C^{zz}_r(t)$ with $h=0.8J$. (b) Long-time behaviour of $\Re[F^{zz}_r(t)]$. Time is rescaled with $\alpha = 0.5066$, so that the curves coincide at long-times and can be compared with the power-law $\sim (Jt)^\beta$, with $\beta = -1.3$.\label{fig: zz OTOC}}
\end{figure}

Below we present results of numerical evaluation of the spin OTOCs for $N=100$ sites. 
Calculations are performed by randomly sampling over charge configurations appearing in Eq.~\eqref{eq: disorder averaged F}, with 20,000 samples in Fig.~\ref{fig: zz OTOC}(b) and 10,000 in all other figures.
%Calculations are performed using Eq.~\eqref{eq: disorder averaged F} where instead of summing over all charge sectors, we use 10,000 randomly chosen configurations in all figures except Fig.~\ref{fig: zz OTOC}(b) where we use 20,000. 
See Ref.~\cite{Smith2017} and the SM for discussion of the convergence of the random sampling. We fix $j$ to be the central bond so that the correlators become functions of the distance $r$ from this bond. We plot data for $h/J = 0.4, 0.8$ where the localization length is sufficiently small compared with system size.

\begin{figure}[t]
	\centering
	\subfigimg[width=.42\textwidth]{\hspace*{0pt} \textbf{}}{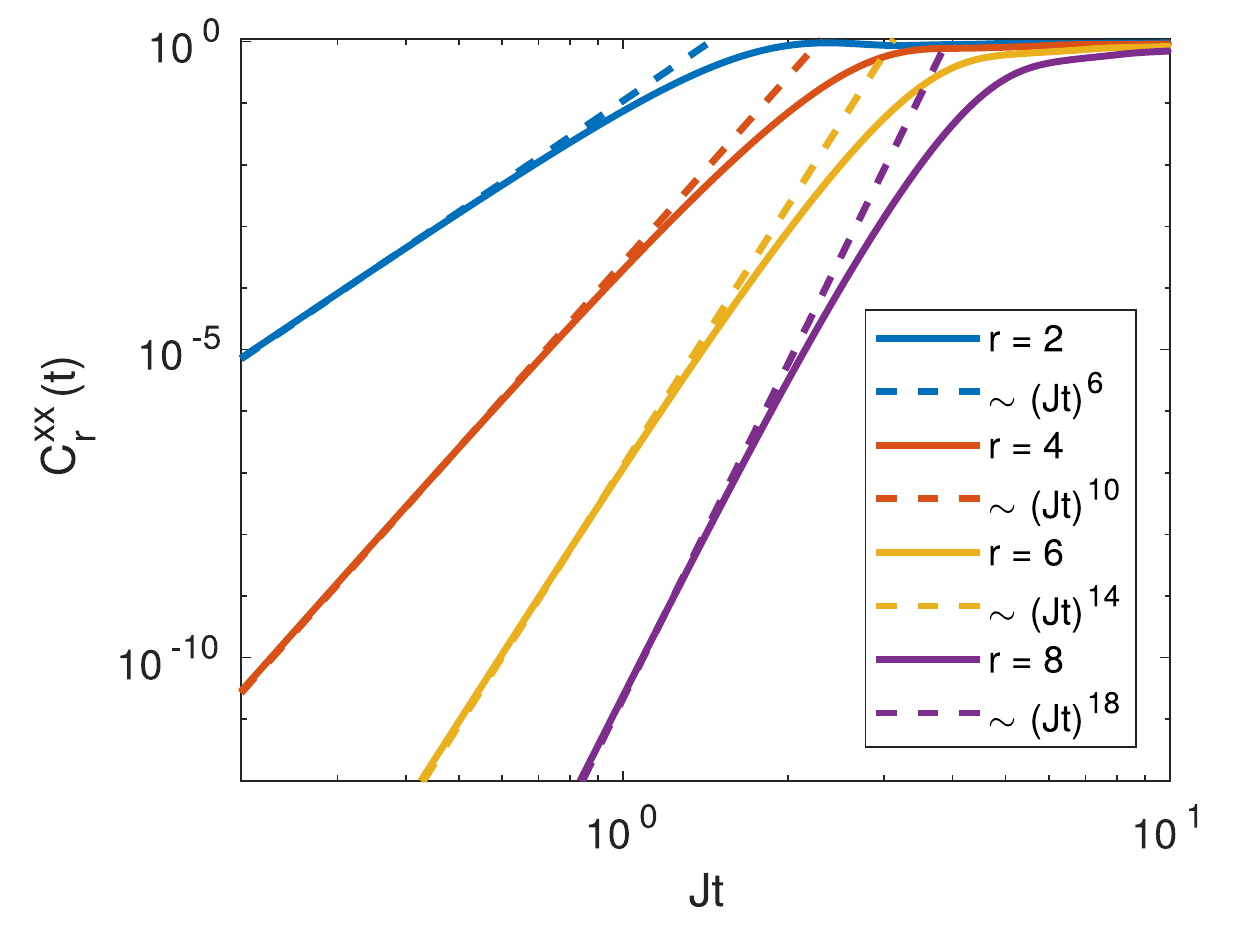}
	\caption{Short-time behaviour of the OTOCs on a log-log scale. Results are shown for $C^{xx}_r(t)$ after a quench with $h = 0.4J$, and we find a similar power-law growth for all cases that we studied. Dashed lines correspond to short-time asymptotics of the form $(Jt)^{2r+2}$.}\label{fig: short time OTOC}
\end{figure}

\textit{Short-Time Behaviour.}~In all cases that we studied (quenches for different components of spin and values of $h$) we observe power-law growth of the OTOC, as shown, e.g.,~in Fig.~\ref{fig: short time OTOC}, consistent with the discussion in Refs.~\cite{Lin2018,Dora2017}. This power-law behaviour can be extracted from the Baker-Campbell-Hausdorff expansion for the time evolution of the operators,
\begin{equation}
\hat{\sigma}^\alpha_{j}(t) = \sum_{n=0}^\infty \frac{(it)^n}{n!} [\hat{H},\hat{\sigma}^\alpha_{j}]_n.
\end{equation}
At leading order the OTOC behaves as $t^{2n}/(n!)^2$, where $n$ is given by the smallest value for which $\big[[\hat{H},\hat{\sigma}^\alpha_{j}]_n,\hat{\sigma}^\beta_{l}\big]$ does not vanish. Since $\hat{H}$ is a local Hamiltonian, the operator $[\hat{H},\hat{\sigma}^\alpha_{j}]_n$ has finite support proportional to $n$, and the lowest-order contribution to the OTOCs arise when $n$ is of the order of the distance between spins $r$. This analysis agrees with the observed short-time behaviour. In particular, we find that $C^{xx}_r(t)$ has the asymptotic form $\sim(Jt)^{2r+2}$, see Fig.~\ref{fig: short time OTOC}.

The authors of Ref.~\cite{Lin2018} suggested that similar arguments hold for any OTOC of bounded local operators whose time evolution is generated by a local Hamiltonian. However, this is in contrast to the exponential growth observed in models with semiclassical limits~\cite{Aleiner2016,Kitaev2015,Maldacena2016a}, and is our second key result.

\begin{figure}[t]
	\centering
	\subfigimg[width=.43\textwidth]{\hspace*{0pt} \textbf{(a)}}{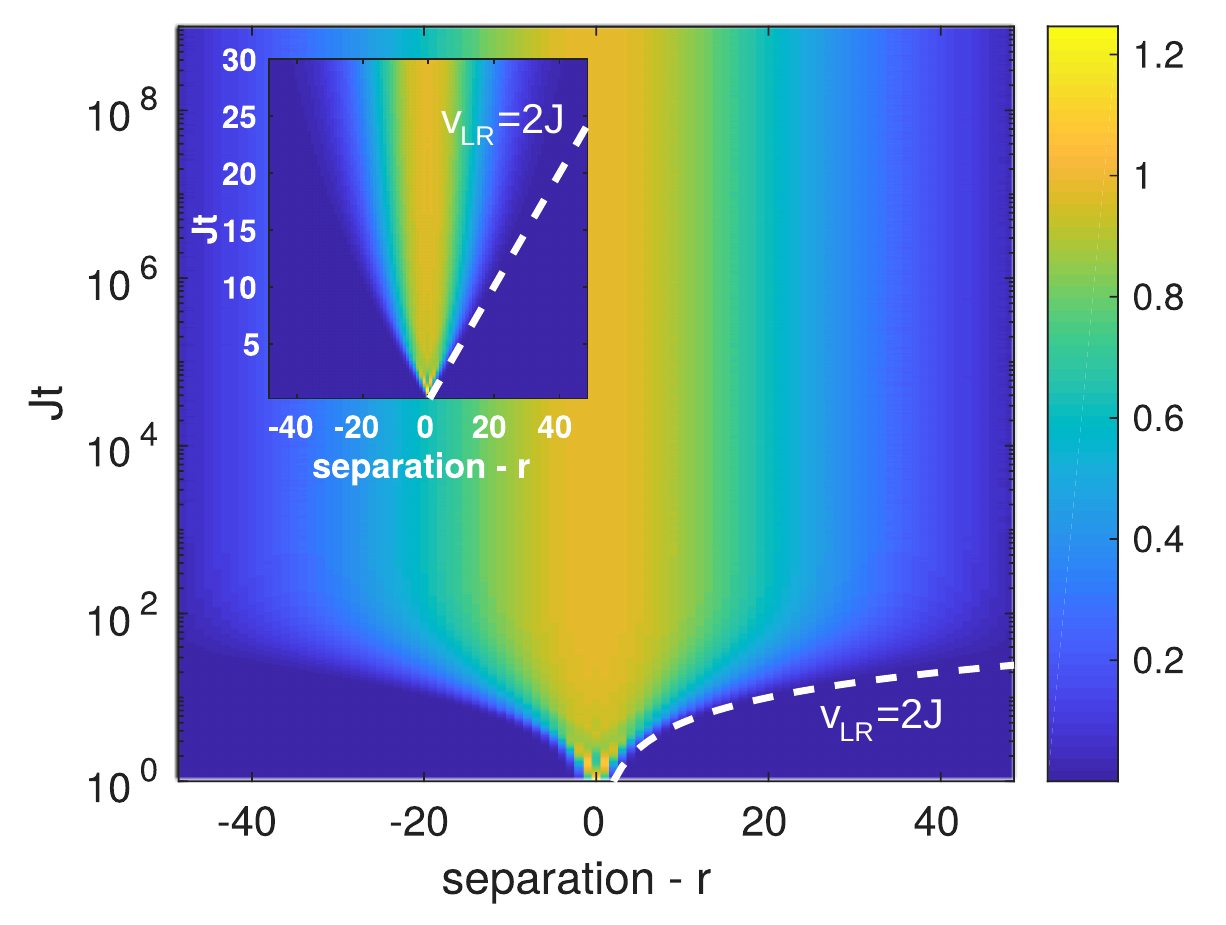}
	\!\!\subfigimg[width=.43\textwidth]{\hspace*{0pt} \textbf{(b)}}{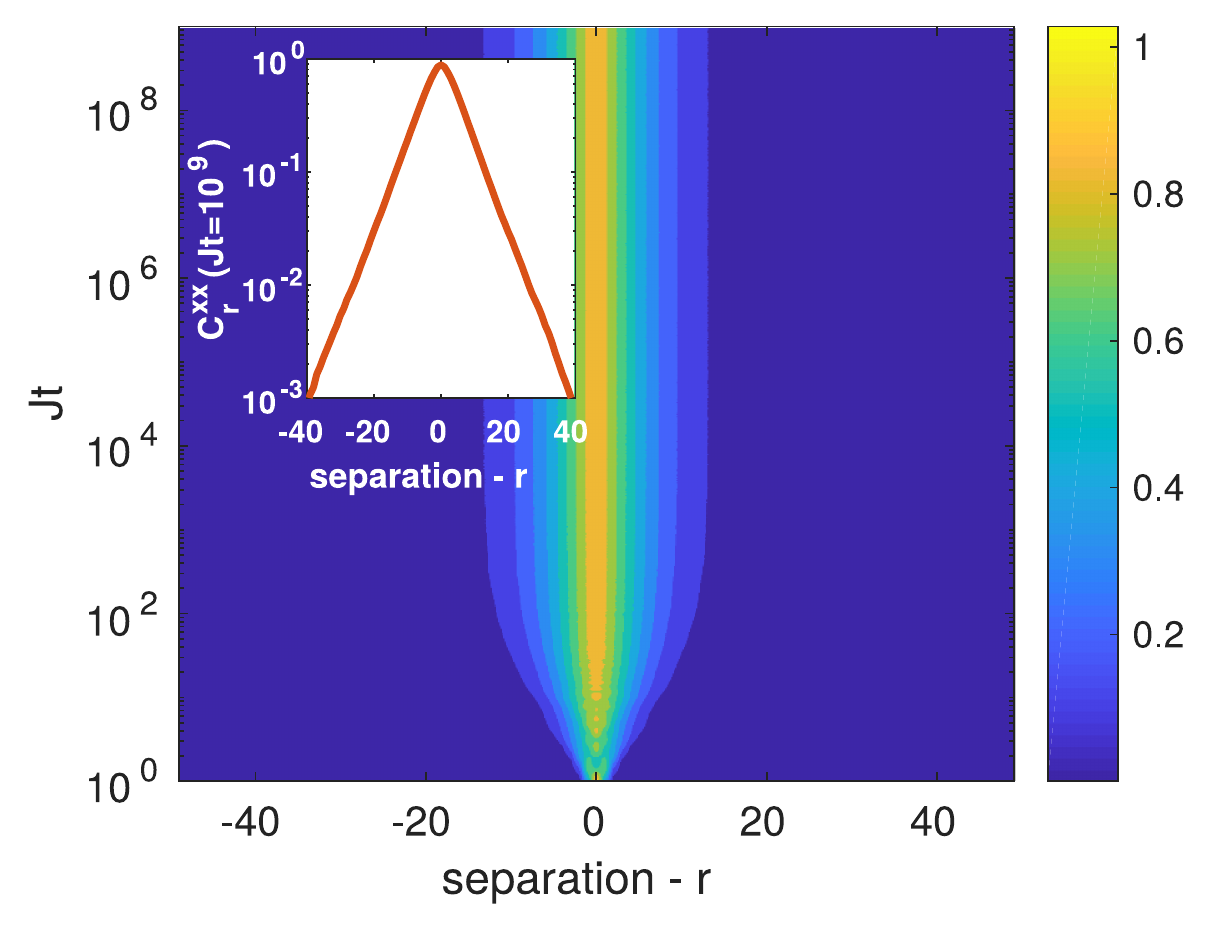}
	\caption{Spreading of the OTOC $C^{xx}_r(t)$ after a quantum quench. (a) Results for $h=0.4J$ on a logarithmic scale. (inset) Short-time behaviour on a linear scale. Dashed line indicates the linear light-cone with Lieb-Robinson velocity $v_\text{LR} = 2J$. (b) Results for $h=0.8J$. (inset) Long time behaviour of correlators on a log-scale.}\label{fig: xx OTOC}
\end{figure}

%We only get a contribution to the OTOC from this expansion when $[\hat{\sigma}^\alpha_{jk}(t),\hat{\sigma}^\beta_{lm}] \neq 0$, in other words when $\big[[\hat{H},\hat{\sigma}^\alpha_{jk}]_n,\hat{\sigma}^\beta_{lm}\big] \neq 0$. Since $\hat{H}$ is a local Hamiltonian, the operator $[\hat{H},\hat{\sigma}^\alpha_{jk}]_n$ must have finite support proportional to $n$, and therefore the lowest order contribution comes when $n$ is proportional to the separation between bonds $\la jk\ra$ and $\la lm\ra$. We can clearly see this short term power-law behaviour in Fig.~\ref{fig: short time OTOC}. The above argument based on the expansion of the time-evolution applies quite generally and in Ref.~\cite{Lin2018} is suggested to hold for any OTOCs of local operators with local Hamiltonian evolution.

\begin{figure}[b]
	\centering
	\!\subfigimg[height=.265\textwidth]{\hspace*{5pt} \textbf{(a)}}{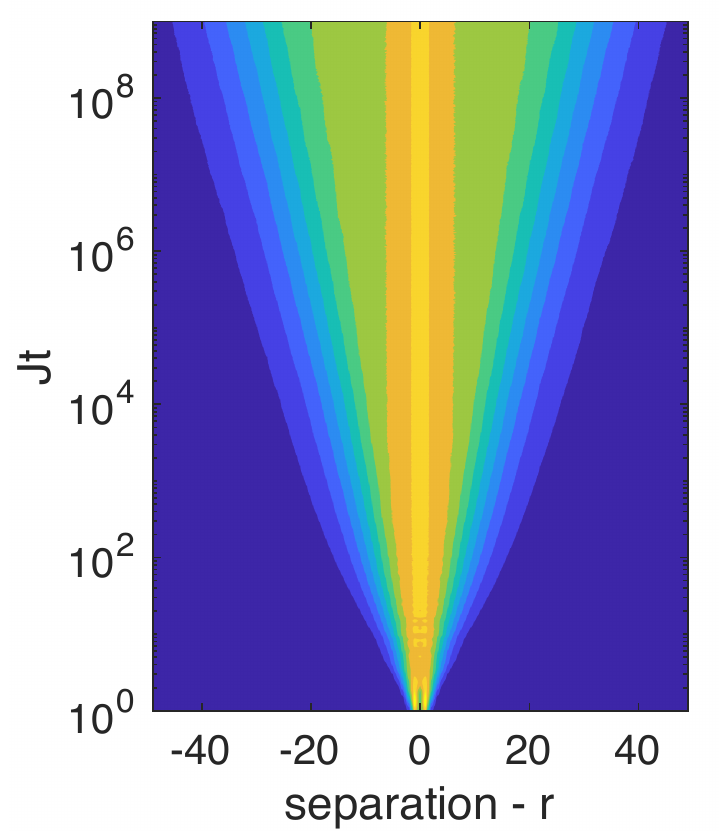}
	\!\!\subfigimg[height=.265\textwidth]{\hspace*{-5pt} \textbf{(b)}}{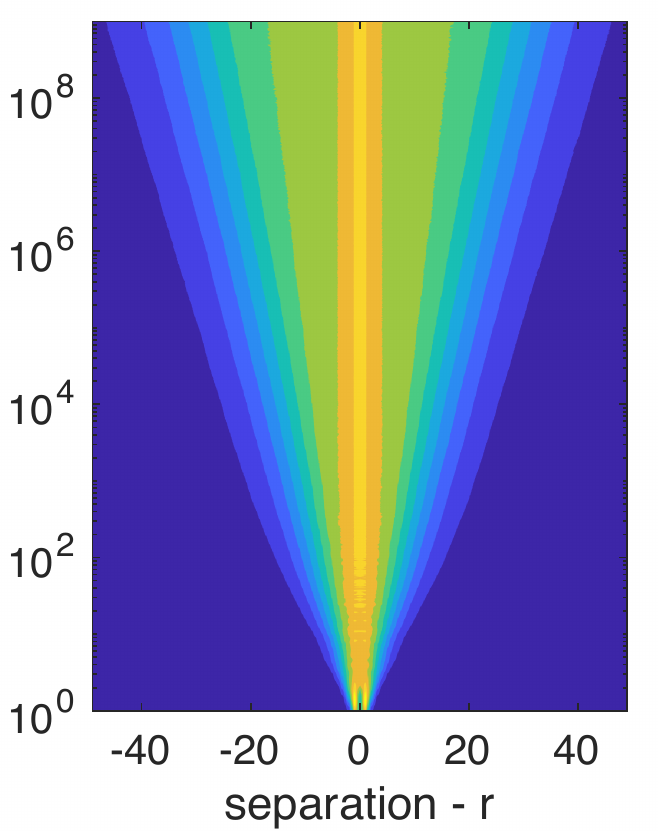}
	\!\!\subfigimg[height=.265\textwidth]{}{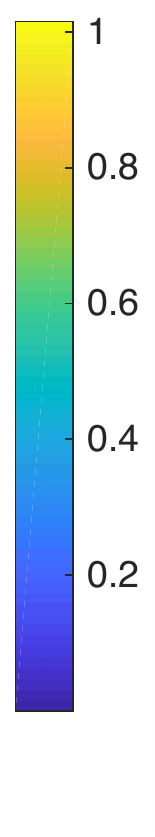}
	\caption{Spreading of the OTOCs involving different spin components: (a) $C^{zx}_r(t)$; (b) $C^{xz}_r(t)$. Results are shown for $h=0.8J$ and on a logarithmic scale.}\label{fig: zx and xz OTOCs}
\end{figure}

\textit{Spreading of correlations.}~In Figs.~\ref{fig: xx OTOC}-\ref{fig: zx and xz OTOCs} we present the correlation spreading in the four distinct spin OTOCs. First, let us discuss the behaviour of $C^{xx}_r(t)$. At short-times, and particularly for small values of the Ising coupling, e.g. $h=0.4J$ shown in Fig.~\ref{fig: xx OTOC}(a), we find a linear light-cone behaviour (see inset), which agrees with the Lieb-Robinson bound with velocity $v = 2J$. At longer times the spreading halts and we find only short-range correlations at long-times. In Fig.~\ref{fig: xx OTOC}(b) we show the spreading with $h=0.8J$ for which the localization length is shorter and the spreading halts more quickly. The inset shows the spatial correlations at long-times, which decay exponentially with separation.

%%The coefficients $\alpha = 0.5066$ and $\beta = -1.3$, are obtained from the fits

This exponential decay of spatial correlations can be understood from Eq.~\eqref{eq: disorder averaged F}. For each charge configuration the Hamiltonians in the forward and backward evolution differ only in the signs of the hopping coefficients $J_j$ on particular bonds. However, as we show in the SM, these changes in sign can be gauged away, i.e. the spectrum of the free-fermion Hamiltonian~\eqref{eq: H free fermions} is independent of the signs of hoppings, and the correlator can only decay via particle transport between the sites $j$ and $l$. However, this transport is exponentially suppressed in the separation $r$ for typical disordered charge configurations due to the Anderson localization of the fermions.

%This exponential decay of spatial correlations can be understood by mapping the OTOC to a disorder-averaged OTOC for free fermions. In the appendix we show that it can be mapped to 
%\begin{equation}
%C_{xx}(t) = \frac{1}{2^{N-1}}\sum_{\{q_j\}\pm 1} \la \psi |[\hat{R}_j(t),\hat{R}'_l]|^2 |\psi \ra,
%\end{equation}
%where $\hat{R}_j(t) = e^{i\hat{H}(q)t} \hat{R}_j e^{-i\hat{H}(q)t}$, and $\hat{R}_j = \prod_{i \leq j} (-1)^{\hat{n}_i}$ and $\hat{R}'_l = \prod_{i > j} (-1)^{\hat{n}_i}$, where we consider $j<l$, without loss of generality. These operators measure the fermion parity on the sites to the left of $j$ and the right of $l$, respectively. Importantly, this commutator can only become non-zero due to particle transport between the left of site $j$ and the right of site $l$. However, for a typical charge configuration we have exponentially localized eigenstates and therefore we have that $C_{xx}(t) \propto e^{-r/\zeta}$ as $t\rightarrow\infty$, where $\zeta$ is proportional to the single-particle localization length. See the appendix for more details.

Next we discuss the logrithmic spreading of the $C^{zz}_r(t)$ correlator, which is one of the main results of our paper. The logrithmic behaviour is evident from the linear form of the contours on a semi-log plot, shown in Fig.~\ref{fig: zz OTOC}(a). Note that for the value $h=0.8J$ shown in this figure, the single-particle localization length for the fermions is $\lambda \approx 2.15$, much smaller than the system size and the scale of correlation spreading.

%Next let us discuss the results for the $C^{zz}_r(t)$ correlator in the case of $h=0.8J$, Fig.~\ref{fig: zz OTOC}(a), which exhibits logarithmic spreading of correlations. Note that for $h=0.8J$, the single-particle localization length for the fermions is $\lambda \approx 2.15$, much smaller than the system size as well as the scale of correlation spreading. This logarithmic spreading of the correlations is one of the main results of our paper.

%This behaviour is similar to that observed in Ref.~\cite{Vardhan2017}, where the authors consider Loschmidt echoes in a localized system. The basic argument that they use is that due to the localized nature of the eigenstates, a local perturbation to the Hamiltonian results in modifications of the eigenstates/energies that are exponentially small in the distance from the quench. This in turn leads to a logarithmic spreading of correlations. 

This logarithmic spreading is a result of the local potential quenches appearing in this correlator (see Eq.~\eqref{eq: disorder averaged F}), which cannot be gauged away -- unlike the bond quenches considered above -- and change the spectra of the Hamiltonians in the forward and backward time evolution.
% so that its time-dependent part can decay due to destructive interference of terms that differ in energy. 
In the SM we use the Lehmann representation to show that this quantity is the sum of terms that decay when $\Delta E t \sim 1$, where $\Delta E=E_\lambda - E_{\lambda_j} + E_{\lambda_{jl}} - E_{\lambda_l}$, with $|\lambda\ra$ a many-body eigenstate, $E_\lambda$ the corresponding energy, and $E_{\lambda_j},E_{\lambda_{jl}}$ are the perturbed energies due to changes in the potential. When $\Delta E \propto e^{-r/\xi}$ decays exponentially with the separation due to localization of the eigenstates~\cite{Serbyn2013}, we obtain the logarithmic spreading of correlations, as seen in Fig.~\ref{fig: zz OTOC}(a). Note that our arguments are independent of presence/absence of interactions, so that this logarithmic spreading is a generic feature of localized systems, see SM for details.

At long times the $C^{zz}_t(t)$ OTOC has power-law behaviour, as can be seen in Fig.~\ref{fig: zz OTOC}(b), which shows the time-dependent piece $F^{zz}_r(t)$. The exponent appears to be approximately independent of the separation as we find by scaling the time by $Jt \rightarrow \alpha^r Jt$ such that the curves coincide with each other. The value $\alpha = 0.5066$ and the exponent $\beta = -1.3$ of the power-law are found empirically. The authors of Refs.~\cite{Vardhan2017,Serbyn2017} also found power-law decay of the Loschmidt echo for localized systems. Similar power-law decay was also observed in the context of OTOCs in a many-body localized system~\cite{Lee2018}, the transverse-field Ising model~\cite{Lin2018}, as well as for the XY spin-chain and symmetric Kitaev chain in Ref.~\cite{McGinley2018}.

Finally, we consider two inequivalent OTOCs involving different spin components $\hat{\sigma}^z$ and $\hat{\sigma}^x$, namely $C^{zx}_r(t)$ and $C^{xz}_r(t)$, see Fig.~\ref{fig: zx and xz OTOCs}. These correlators show qualitatively similar behaviour. For short separations we find nearly time independent contours signifying localization behaviour. However, for larger separations we observe additional spreading of correlations. While the analytic arguments presented above do not apply to this case of mixed-component correlators, the logarithmic spreading appears to be a more general feature.

\textit{Discussion.}~We have studied four distinct OTOCs for the spins in a model of disorder-free localization. These present a remarkably rich phenomenology beyond that which has been observed for Anderson-localized systems. We show that OTOCs in our model can be mapped onto a disorder-averaged double Loschmidt echo. Perhaps unsurprisingly, we do not find exponential growth of the OTOC that has been attributed to chaotic behaviour. Instead we find short-time power-law growth consistent with that found for other integrable models~\cite{Lin2018,Dora2017,McGinley2018}. 
While our model does not contain the ingredients of many-body localization, we find correlation spreading which is logarithmic, and in some cases the model shows a complete lack of correlation spreading. 

We suggest that the logarithmic spreading arises as a result of the double Loschmidt echo form of the spin correlators, which are unlike the usual correlators appearing in fermion models with quenched disorder. When the local perturbations in the double Loschmidt echo correlator change the energy spectrum we get logarithmic spreading. We note that a similar slow spreading of free-fermion OTOCs has been observed in Ref.~\cite{McGinley2018} due to a different mechanism, namely a non-local form of the operators in the computational basis. 
%Logarithmic entanglement growth has also been in observed in the fine-tuned critical phase of a non-interacting central-site model~\cite{Hetterich2017}, and a logarithmic correction to the inverse participation ratio was derived for a free-fermion model in Ref.~\cite{Arias1998}.
A logarithmic correction to the inverse participation ratio was also derived for a free-fermion model in Ref.~\cite{Arias1998}, and logarithmic entanglement growth was observed in the critical phase of a non-interacting central-site model~\cite{Hetterich2017} 
In cases where changes of bond signs can be gauged away, we find exponentially localized correlations, similar to the standard fermion correlators in quenched disorder models of Anderson localization, see e.g. Ref.~\cite{Huang2017}.

Although these quantities arise as natural spin OTOCs in our model, we propose that the resulting double Loschmidt echo form of the correlators may be useful in the studies of correlation spreading more generally. We stress that the analytical arguments for logarithmic behaviour do not depend on specific features of our model and apply more generally to other localized systems. Our work shows that the spreading of correlations, operators and perturbations under unitary evolution -- as probed by standard correlation functions, OTOCs and the double Loschmidt echo, respectively -- can all be distinct. Additional tools, such as the double Loschmidt echo may therefore be required to fully characterize the possible non-equilibrium behaviour.  

Our model provides an ideal setting for further studies of OTOC phenomenologies because of a free-fermion mapping which allows one to access large system sizes and gain analytical insights, which can also be used for thermal or infinite-temperature OTOCs. We leave a full investigation of the initial state and temperature dependence of the OTOCs to future work. It is worth stressing that the gauge-invariance of the spectrum under bond quenches is not a special feature of our model, but applies more generally to, e.g.,~a $\mathbb{Z}_2$ representation of the Hubbard model~\cite{Smith2018}. Unfortunately, in these cases one is facing severe computational limitations on system sizes and time scales. Remarkably, there are also prospects of simulating OTOCs in experiments~\cite{Yao2016,Garttner2017}. These precisely controlled systems may provide access to strongly-correlated physics beyond numerical capabilities.

\begin{acknowledgements}\textit{Acknowledgements.}~We thank F.~Pollmann, A.~Nahum, G.~De Tomasi and K.~H{\'e}mery for discussions. A.S.~acknowledges EPSRC for studentship funding under Grant No.~EP/M508007/1. The work of D.K.~was supported by EPSRC Grant No.~EP/M007928/2, R.M.\ was in part supported by DFG under grant SFB 1143 and EXC 2147.
\end{acknowledgements}

\appendix

\hfill\\
\noindent\makebox[\linewidth]{\resizebox{0.7\linewidth}{1pt}{$\bullet$}}\bigskip

\begin{center}
	\textbf{Supplemental Material}
\end{center}

In this supplemental material we provide explicit details of the arguments used in the main text to understand the observed behaviour for the $C^{xx}_r(t)$ and $C^{zz}_r(t)$ OTOCs, an interpretation of the double Loschmidt protocol, and evidence for the convergence of the random charge configuration sampling used in the main text. Please see the appendices of Refs.~\cite{Smith2017,Smith2018} for details of the determinant method used for our numerical simulations.

\emph{Exponential localization of $C_r^{xx}(t)$.}
The time-dependent part of the $C_r^{xx}(t)$ OTOC is 
\begin{equation}\label{eq: Fxx appendix}
F_r^{xx}(t) = \frac{1}{2^{N-2}} \sum_{\{q_i\} = \pm1} \la \psi | e^{i\hat{H}(q)t} e^{-i\hat{H}^{x}_{j}(q)t} e^{i \hat{H}^{xx}_{jl}(q)t} e^{-i\hat{H}^{x}_{l}(q)t} | \psi\ra,
\end{equation}
where for $\hat{H}^x_j(q)$ we have $J_{j} \rightarrow - J_{j}$, for $\hat{H}^{xx}_{jl}(q)$ we have $J_{j},J_{l} \rightarrow - J_{j},- J_{l}$, and for $\hat{H}^x_l(q)$ we have $J_{l} \rightarrow - J_{l}$, all relative to $\hat{H}(q)$. These four Hamiltonians differ locally by the sign of the relevant tunnelling parameter, i.e., local bond quenches. We proceed by noting that the transformation $J_{j} \rightarrow -J_{j}$ is equivalent to $\hat{c}_i \rightarrow -\hat{c}_i$, for $i\leq j$. This transformation can be implemented by the unitary string operator $\hat{R}_j = \prod_{i \leq j} (-1)^{\hat{n}_i}$ or equivalently by $\hat{R}'_j = \prod_{i > j} (-1)^{\hat{n}_i}$. For instance, $e^{\pm i\hat{H}^{x}_{j}(q)t} = \hat{R}_j e^{\pm i\hat{H}(q)t} \hat{R}_j$. Importantly, this means that the spectrum is unchanged by the local bond quenches.

Let us, without loss of generality, consider $l > j$, then using the string operators we can rewrite the time-dependent piece of the OTOC~\eqref{eq: OTOC F} as 
\begin{equation}
F_r^{xx}(t) = \frac{1}{2^{N-2}}\sum_{\{q_j\}\pm 1} \la  \psi | e^{i\hat{H}(q)t} \hat{R}_j e^{-i\hat{H}(q)t} \hat{R}'_l e^{i\hat{H}(q)t} \hat{R}_j e^{-i\hat{H}(q)t} \hat{R}'_l |\psi\ra,
\end{equation}
where we have used the facts that $[\hat{R}_j,\hat{R}'_l] = 0$ and $(\hat{R}_j)^2 = 1$. The further means that the OTOC can be written as
\begin{equation}\label{eq: string OTOC}
C_r^{xx}(t) = \frac{1}{2^{N-1}}\sum_{\{q_j\}\pm 1} \la \psi ||[\hat{R}_j(t),\hat{R}'_l]|^2 |\psi \ra,
\end{equation}
where $\hat{R}_j(t) = e^{i\hat{H}(q)t} \hat{R}_j e^{-i\hat{H}(q)t}$, which implicitly depend on the charge configuration. Note the extra prefactor of $1/2$ coming from the definition of the OTOC in Eq.~\eqref{eq: OTOC def}. This OTOC is therefore reduced to a disorder-averaged OTOC of parity operators for the free fermion Hamiltonian~\eqref{eq: H free fermions}. We can understand the observed behaviour by considering the operator commutator for random configurations of the potential, i.e., for the individual terms of the sum in Eq.~\eqref{eq: string OTOC}.

The operator $\hat{R}_j$ measures the fermion parity on the chain of sites to the left of $j$, and similarly, $\hat{R}'_l$ measures the parity to the right of site $l$. Therefore, to get $[\hat{R}_j(t),\hat{R}'_l] \neq 0$ would require particle transport between the left of site $j$ and the right of site $l$ due to the time evolution. However, since for a typical charge configuration we have a disordered potential, the fermions will be localized meaning that there is an exponentially small probability for the required particle transport. This implies that $\la \psi ||[\hat{R}_j(t),\hat{R}'_l]|^2|\psi\ra \propto e^{-|j-k|/\zeta}$ as $t\rightarrow \infty$, where $\zeta$ is a length scale which is a function of the single-particle localization lengths. This long-time behaviour is observed in Fig.~\ref{fig: xx OTOC}(b), where we find exponential tails for the spatial distribution of correlations.

The arguments used here are not specific to our model, and should apply to correlators of the form in Eq.~\eqref{eq: Fxx appendix} for more general Anderson or many-body localized systems. The important ingredients are that the bond perturbations do not change the spectrum and that particle transport is exponentially suppressed with distance.

\emph{Logarithmic spreading of $C_{zz}(t)$.}
Whereas $C_r^{xx}(t)$ corresponded to a disorder-averaged double Loschmidt echo procedure with local bond quenches, the quenches in the $C_r^{zz}(t)$ OTOC are of the local potential. Importantly, this type of quench does not preserve the spectrum. The time-dependent part of the $C_r^{zz}(t)$ OTOC is 
\begin{equation}
F_r^{zz}(t) = \frac{1}{2^{N-2}} \sum_{\{q_i\} = \pm1} \la \psi | e^{i\hat{H}(q)t} e^{-i\hat{H}^{z}_{j}(q)t} e^{i \hat{H}^{zz}_{jl}(q)t} e^{-i\hat{H}^{z}_{l}(q)t} | \psi\ra,
\end{equation}
where for $\hat{H}^z_j(q)$ we have $h_{j},h_{j+1} \rightarrow - h_{j},-h_{j+1}$, for $\hat{H}^{zz}_{jl}(q)$ we have $h_{j}, h_{j+1},h_{l},h_{l+1} \rightarrow -h_{j}, -h_{j+1},-h_{l},-h_{l+1} $, and for $\hat{H}^z_l(q)$ we have $h_{l}, h_{l+1} \rightarrow - h_{l}, h_{l+1}$, all relative to $\hat{H}(q)$. This correlator corresponds to a double Loschmidt echo procedure with local density quenches of the form $\hat{V} = \pm 2h(\hat{n}_j \pm \hat{n}_{j+1})$, where the signs depend on the particular charge configuration.

We can understand the logarithmic spreading seen in Fig.~\ref{fig: zz OTOC}(a) using perturbative arguments similar to those in Ref.~\cite{Vardhan2017} for the long-time behaviour of the standard Loschmidt echo. More explicitly, let us consider the single-particle correlators
\begin{equation}
\mathcal{L}_r^{zz}(t) = \la \psi | e^{i\hat{H}(q)t} e^{-i\hat{H}^{z}_{j}(q)t} e^{i \hat{H}^{zz}_{jl}(q)t} e^{-i\hat{H}^{z}_{m}(q)t} | \psi\ra, 
\end{equation}
for a typical charge configuration, which are averaged over in Eq.~\eqref{eq: OTOC F}. We can use the Lehmann representation to express it in terms of the many-body eigenstates and energies. This gives
\begin{equation}\label{eq: L Lehmann 1}
\mathcal{L}_r^{zz}(t) =\!\sum_{\substack{\lambda,\mu_{j},\nu_{jl},\chi_{l}}} \!\la \psi | \lambda \ra \la \lambda | \mu_{j} \ra \la \mu_{j} | \nu_{jl} \ra \la \nu_{jl} | \chi_l \ra \la \chi_l |\psi\ra e^{i\Delta Et},
\end{equation}
where $\Delta E = E_\lambda - E_{\mu_j} + E_{\nu_{jl}} - E_{\chi_{l}}$. In this expansion, the states $|\lambda\ra,|\mu_j\ra,|\nu_{jl}\ra, |\chi_{l}\ra$ are many-body eigenstates of the Hamiltonians $\hat{H}(q),\hat{H}^z_{j}(q),\hat{H}^{zz}_{jl}(q),\hat{H}^{z}_{l}(q)$, respectively, and $E_{\lambda},E_{\mu_j},E_{\nu_{jl}},E_{\chi_l}$ are the corresponding energy eigenvalues.

We then proceed by first making the approximation that the wavefunctions are only locally perturbed and, in particular, we assume that $\la \lambda | \mu_j \ra \approx \delta_{\lambda,\mu_j}$, and similarly for all the eigenstate overlaps appearing in Eq.~\eqref{eq: L Lehmann 1}, and we neglect the modification of the eigenvectors due to the perturbation. This is justified by the fact that the single-particle eigenstates are localized and the local potential perturbation only locally changes the eigenstates, and in particular the exponential profile is preserved. We will discuss the validity of this assumption below, see also Refs.~\cite{Cerruti2002,Vardhan2017,Serbyn2013}.

With this approximation taken into account, the expression~\eqref{eq: L Lehmann 1} reduces to 
\begin{equation}\label{eq: L Lehmann}
\mathcal{L}_r^{zz}(t) \approx \sum_\lambda \la \psi | \lambda \ra \la \lambda | \psi \ra e^{i\Delta E(\lambda)t},
\end{equation}
where $\Delta E(\lambda) = E_\lambda - E_{\lambda_j} + E_{\lambda_{jl}} - E_{\lambda_{l}}$. The deviation of $\mathcal{L}_r^{zz}(t)$ from $1$ is thus determined by $\Delta E(\lambda)$, wherein the energies $E_{\lambda_j}, E_{\lambda_{jl}}, E_{\lambda_l}$ are the perturbed energies corresponding to the same eigenstate $|\lambda\ra$. We estimate this energy difference using a second-order perturbation expansion.

For convenience, let us state the result of perturbation theory for the eigenergies up to second order (see, e.g., Ref.~\cite{Landau_QM}). Consider a Hamiltonian $\hat{H}$ and a perturbation $\epsilon \hat{V}$. In our case, $\hat{V}$ takes the form $\hat{V}_{j} = \pm \hat{n}_j \pm \hat{n}_{j+1}$, $\hat{V}_{jl} = \pm \hat{n}_j \pm \hat{n}_{j+1} \pm \hat{n}_l \pm \hat{n}_{l+1}$, or $\hat{V}_{l} = \pm \hat{n}_l \pm \hat{n}_{l+1}$, with $\epsilon = 2h$. The energies of the perturbed Hamiltonian $\hat{H} + \hat{V}$ are then given by
\begin{equation}
E_\lambda = E^{(0)}_\lambda + \epsilon \la \lambda| \hat{V} |\lambda \ra + \epsilon^2 \sum_{\mu \neq \lambda} \frac{|\la\lambda| \hat{V} |\mu \ra|^2} {E^{(0)}_\lambda - E^{(0)}_\mu} + \mathcal{O}(\epsilon^3),
\end{equation}
where $|\lambda\ra,|\mu\ra$ are unperturbed eigenstates, and $E^{(0)}_\lambda,E^{(0)}_\mu$ are the unperturbed energy eigenvalues.
The first-order corrections to the energies cancel in the energy difference $\Delta E(\lambda)$ and the leading-order correction from second-order is
\begin{equation}\label{eq: second order correction}
\Delta E(\lambda) \approx 
\pm 8h^2 \text{Re}\sum_{\mu \neq \lambda} \frac{\la \lambda| \hat{n}_j \pm \hat{n}_{j+1} | \mu \ra \la \mu | \hat{n}_l \pm \hat{n}_{l+1} | \lambda \ra } {E^{(0)}_\lambda - E^{(0)}_\mu},
\end{equation}
where the signs are determined by the particular charge configuration. Since for a typical disorder configuration the system is Anderson localized, the energy eigenvalues are uncorrelated and the difference $E^{(0)}_\lambda - E^{(0)}_\mu$ is a random function of the states $|\lambda\ra,|\mu\ra$.
The operator 
\begin{equation}
\sum_{\mu \neq \lambda} \frac{ | \mu \ra \la \mu |} {E^{(0)}_\lambda - E^{(0)}_\mu},
\end{equation}
is therefore a random diagonal operator in the basis of eigenstates. Furthermore, for a given localized eigenstate $|\lambda\ra$, the density correlations in Eq.~\eqref{eq: second order correction} are exponentially decaying with the separation $r$. The energy difference to second-order will therefore take the functional form
\begin{equation}
\Delta E(\lambda) \sim \pm 8h^2C(\{E_\mu\},|\lambda\ra)\; e^{-r/\xi}, 
\end{equation}
where $C(\{E_\mu\},|\lambda\ra)$ is a random function of the energies of the many-body states of $\hat{H}(q)$ with dimensions of inverse energy, $r$ is the separation between the bonds $j$ and $l$, and the length scale $\xi$ is a function of the single-particle localization lengths. Plugging this back into Eq.~\eqref{eq: L Lehmann} we find that the correlator deviates from $1$ when $h^2 C e^{-r/\xi}t \sim 1$. Rearranging this expression we find the relationship 
\begin{equation}
r \sim \ln\left(\frac{h^2 C t}{\xi}\right),
\end{equation}
which defines the light-cone for the OTOC. Averaging over $|\lambda\ra$ and the charge configurations $\{q_j\}$ results in a logarithmic light-cone, as observed in Fig.~\ref{fig: zz OTOC}(a).

While these arguments are perturbative, we also see this logarithmic light-cone beyond the perturbative regime. We argue that they can be extended due to the decreasing localization length with increasing $h$. This means that the orthogonality assumptions, e.g., $\la \lambda | \mu_j \ra \approx \delta_{\lambda,\mu_j}$, remain valid. The energy corrections to all orders are also correlators of density operators and random diagonal operators and therefore decay exponentially with the separation $r$. 

So long as the assumption, $\la \lambda | \mu_j \ra \approx \delta_{\lambda,\mu_j}$, remains approximately valid the above arguments should also apply for more general Anderson and many-body localized systems. This assumption should remain valid when we have quasi-local conserved $l$-bits and the strength of the local perturbation is sufficiently small. Therefore, the observed logarithmic spreading of the double Loschmidt echo with density perturbations is a generic feature of localized systems, with or without interactions.

\begin{figure}[t!]
	\centering
	\includegraphics[width=.43\textwidth]{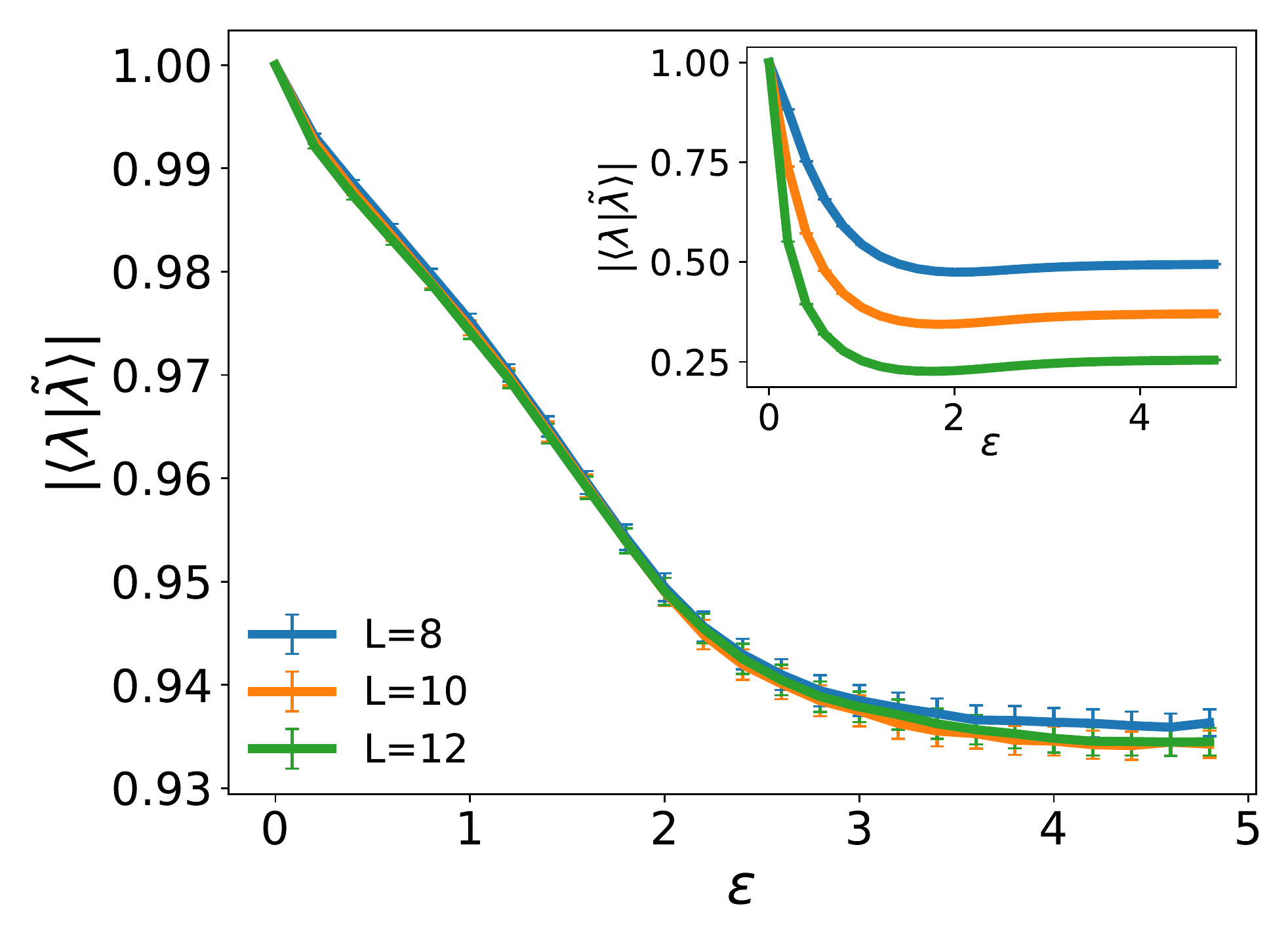}
	\caption{Average overlap between a state $|\lambda\ra$ and the corresponding eigenstate $|\tilde{\lambda}\ra$ of the perturbed Hamiltonian. The data is for the Heisenberg chain with disorder strength $W=10$ in the main figure and $W=0.5$ in the inset, plotted against the perturbation strength $\epsilon$. The overlap is averaged over eigenstates 2000 disorder realizations and the error bars correspond to the standard error of the mean. }\label{fig: eigenstate overlap}
\end{figure}

In Fig.\ref{fig: eigenstate overlap} we include some numerical evidence that suggests that this approximation is indeed valid for MBL systems. We consider the isotropic Heisenberg spin-chain with disorder strength $W=10$ in units of the coupling and a density perturbation on the central site of strength $\epsilon$. This data shows the average value of $|\la \lambda | \tilde{\lambda} \ra|$ -- where $|\tilde{\lambda}\ra$ is the perturbed eigenstate corresponding to $|\lambda\ra$ -- as a function of the perturbation strength, $\epsilon$. This shows that for small perturbations this overlap is strongly dominant and approximately independent of system size. This also seems to hold beyond the small perturbation regime, as we also found numerically for our model, and is in stark contrast to the behaviour for states in the ergodic phase for $W=0.5$ shown in the inset.

\emph{Interpretation of the double Loschmidt echo.}	
Here we present an interpretation of the double Loschmidt echo, which also further validates the numerical results of the main text. The time dependent part of the double Loschmidt echo is of the form
\begin{equation}\label{eq: OTOC F appendix}
F^{\alpha\beta}_{jl}(t) = \la \Psi | e^{i\hat{H}t} e^{-i\hat{H}^\alpha_{j}t} e^{i \hat{H}^{\alpha\beta}_{jl}t} e^{-i\hat{H}^\beta_{l}t} | \Psi\ra,
\end{equation}
where here we slightly generalize the form found in Eq.~\eqref{eq: OTOC F} found in the main text. We consider $\hat{H}^\alpha_j = \hat{H} + \hat{V}^\alpha_j$, where $\hat{H}^{\alpha\beta}_{jl} = \hat{H} + \hat{V}^\alpha_j + \hat{V}^\beta_l$, where $\hat{V}^\alpha_j$ and $\hat{V}^\beta_j$ are two type of local perturbations labelled by $\alpha$ and $\beta$ and localized on site $j$.

\begin{figure}[b!]
	\centering
	\subfigimg[width=.43\textwidth]{\hspace*{0pt} \textbf{(a)}}{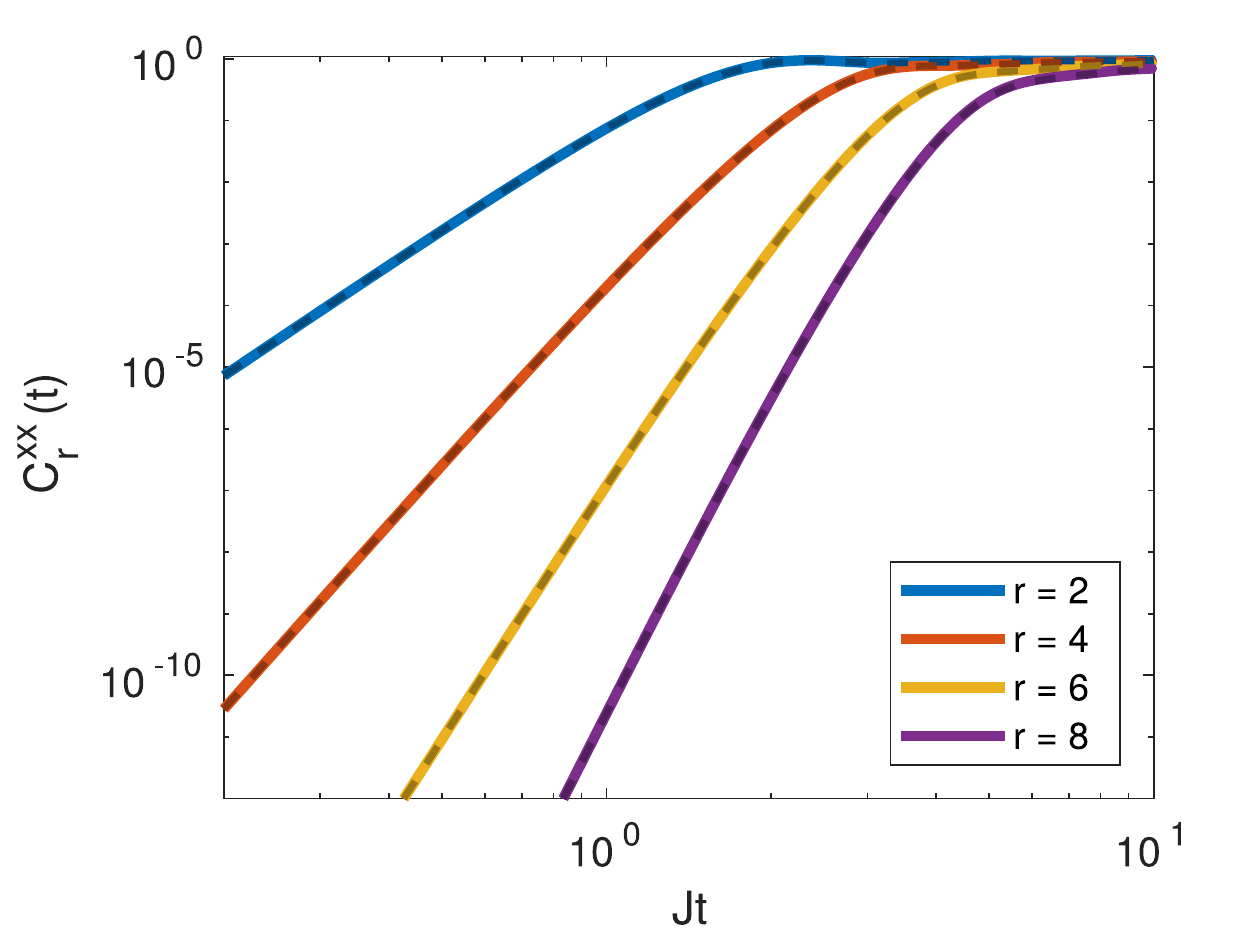}
	\!\!\subfigimg[width=.43\textwidth]{\hspace*{0pt} \textbf{(b)}}{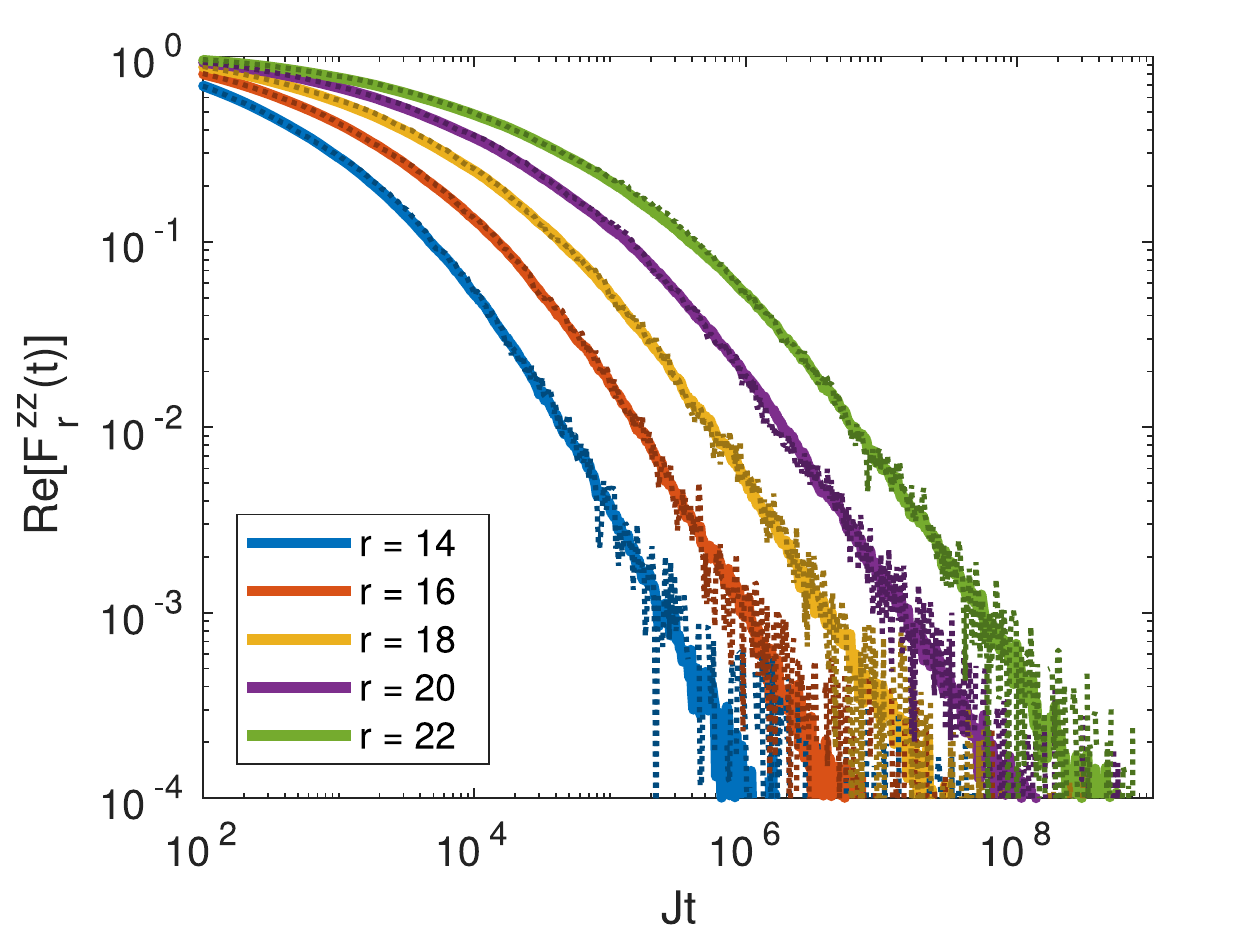}
	\caption{(a) Short-time behaviour of $C^{xx}_r(t)$ with $h=0.4J$ for different values of the separation $r$, comparing averaging over 10,000 (light solid) with 500 (dark dashed). (b) Long time behaviour of $\text{Re}[F^{zz}_r(t)]$ with $h=0.8J$ comparing 20,000 configurations (light solid) with 500 (dark dotted).}\label{fig: short and long time comparison}
\end{figure}

\begin{figure}[b]
	\centering
	\subfigimg[width=.43\textwidth]{\hspace*{0pt} \textbf{(a)}}{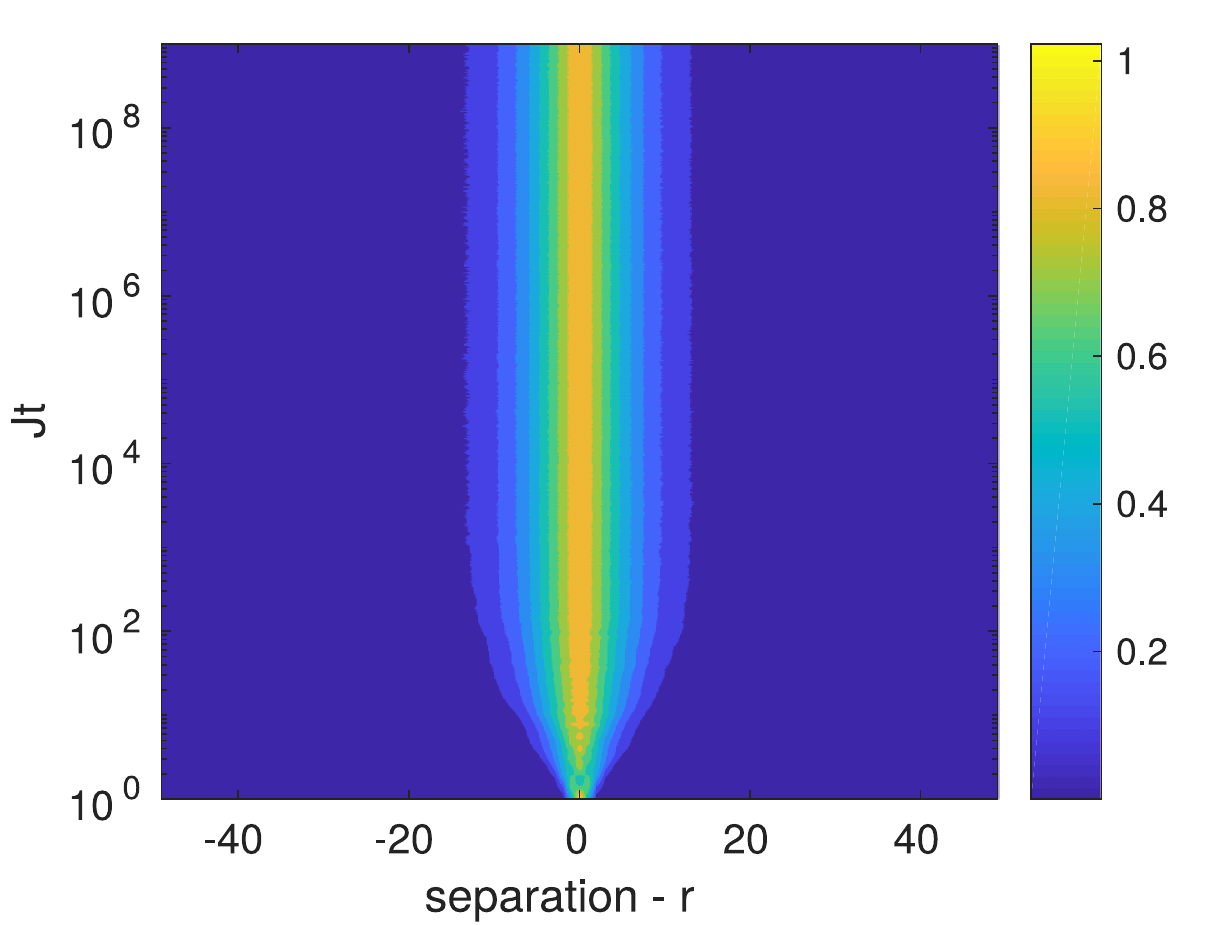}
	\!\!\subfigimg[width=.43\textwidth]{\hspace*{0pt} \textbf{(b)}}{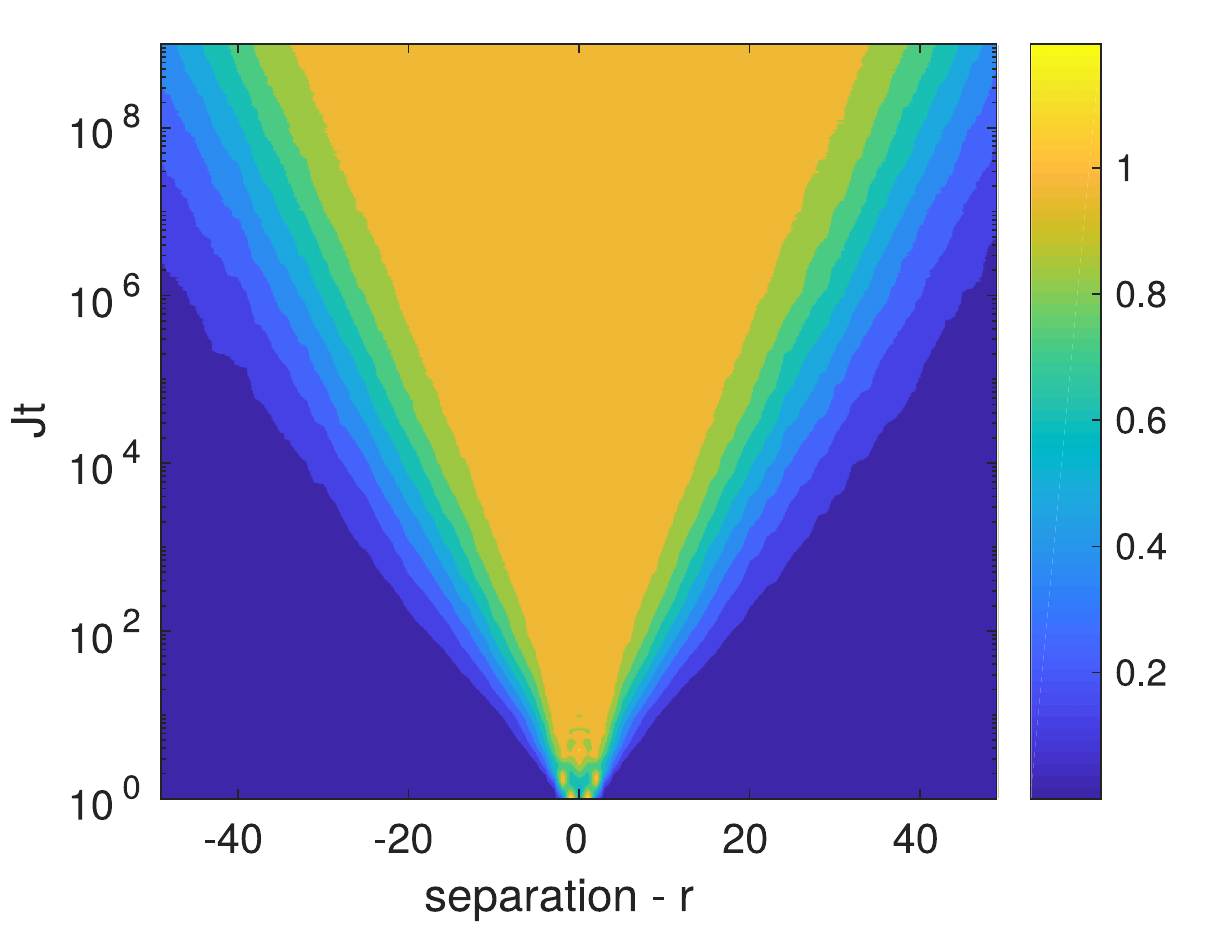}
	\caption{Spreading of OTOCs: (a) $C^{xx}_r(t)$, (b) $C^{zz}_r(t)$. Both figures are for $h=0.8J$ and averaging over 500 charge configurations. To be compared with Fig.~\ref{fig: xx OTOC}(b) and Fig.~\ref{fig: zz OTOC}(a) of the main text, respectively.}\label{fig: xx and zz comparison}
\end{figure}

As a starting point let us consider the two cases where one of the local perturbations is zero. First, if $\hat{V}^\alpha_j = 0$, then we have that
\begin{equation}
F^{\alpha\beta}_{jl}(t)  = \la \Psi | e^{i\hat{H}t} e^{-i\hat{H}t} e^{i \hat{H}^{\beta}_{l}t} e^{-i\hat{H}^\beta_{l}t} | \Psi\ra = 1,
\end{equation}
since each time evolution is followed by its exact time-reversal. Secondly, if instead $\hat{V}^\beta_l = 0$, then
\begin{equation}
F^{\alpha\beta}_{jl}(t)  = | e^{i \hat{H}^{\alpha}_{j}t} e^{-i\hat{H}t} | \Psi\ra |^2 = 1,
\end{equation}
which follows from the unitarity of the Hamiltonian evolution. A non-trivial response is therefore the result of the mutual influence of the two perturbations. Put another way, the response can only be non-trivial if the influence of the local perturbation at site $j$ in 
\begin{equation}
e^{i\hat{H}^\alpha_j t} e^{-i\hat{H} t} |\Psi \ra , \quad \text{and} \quad e^{i\hat{H}^{\alpha\beta}_{jl} t} e^{-i\hat{H}^\beta_l t} |\Psi\ra,
\end{equation}
is influenced by the perturbation at site $l$. The double Loschmidt echo therefore quantifies only the mutual influence of the two separated local perturbations and not the effect of either individually. We hence suggest that this is a natural quantity to consider beyond the scope considered in the main text. In particular, if the perturbations $\alpha$ and $\beta$ are the same type, the double Loschmidt echo measures the spatial influence of that type of perturbation as a function of time and provides more information than the Loschmidt echo alone.

Note, in case where the types of perturbation $\alpha$ and $\beta$ differ, that while swapping the two perturbations in Eq.~\eqref{eq: OTOC F appendix} results in an inequivalent quantity, a non-trivial response in one implies it in the other. This is observed in Fig.~\ref{fig: zx and xz OTOCs} of the main text -- while the two correlators differ in their details, they agree qualitatively.

\emph{Convergence of random sampling.}	
For our numerical results presented in the main text we do not average over all of the $2^{N-2}$ charge configurations appearing in Eq.~\eqref{eq: disorder averaged F}. Instead we average over a randomly sampled set of 10,000--20,000 configurations. This set is only a small percentage of the possible number, but here we demonstrate that it is sufficient for the convergence of our results by comparing with averaging over only 500--1,000. The accuracy of random sampling was also demonstrated in Ref.~\cite{Smith2017}.

In Fig.~\ref{fig: short and long time comparison} we compare data for the short and long time behaviour of the OTOCs for different numbers of samples. We compare the short time behaviour if $C^{xx}_r(t)$ for $h = 0.4J$ (corresponding to Fig.~\ref{fig: short time OTOC} of the main text) using 10,000 configurations against 500 configurations in Fig.~\ref{fig: short and long time comparison}(a) and in Fig.~\ref{fig: short and long time comparison}(b) we compare the long-time behaviour for 20,000 and 1,000 configurations (corresponding to Fig.~\ref{fig: zz OTOC}(b) of the main text).

The short-time behaviour in Fig.~\ref{fig: short and long time comparison}(a) shows no discernable difference between the different number of configurations averaged over. At long-times, shown in Fig.~\ref{fig: short and long time comparison}(b), the different effective disorder configurations have a much larger effect which shows up as random noise when too few samples are taken. However, even for 1,000 samples the correct overall power-law decay is still visible. For 20,000 samples the fluctuations are strongly suppressed.

In Figs.~\ref{fig: xx and zz comparison}(a-b) we show the $C^{xx}_r(t)$ and $C^{zz}_r(t)$, respectively, for $h=0.8J$ averaging over only 500 random charge configurations. These figures should be compared with Fig.~\ref{fig: xx OTOC}(b) and Fig.~\ref{fig: zz OTOC}(a) of the main text, which used 10,000 configurations. These figures show a clear qualitative agreement with those averaged over 10,000, but with additional random fluctuations visible in the contours. Importantly, despite these fluctuations, the correlations in Fig.~\ref{fig: xx and zz comparison}(a) are clearly spatially localized, and spread logarithmically in Fig.~\ref{fig: xx and zz comparison}(b).

These results show that even with as few as 500 randomly samples configurations, the data is still consistent with the conclusions of the main text. By sampling over 10,000-20,000 configurations we are able to remove the visible fluctuations in the data indicating that it has converged to a sufficient accuracy.

\end{document}